\documentclass[prl,twocolumn,superscriptaddress,showpacs]{revtex4}

\usepackage{amsmath}
\usepackage{graphicx}

\newcommand{\T}{${\mathcal T}$}

\begin{document}

\title{Induced Time-Reversal Symmetry Breaking Observed in Microwave Billiards}

\author{B.~Dietz}
%\email{dietz@ikp.tu-darmstadt.de}
\affiliation{Institut f{\"u}r Kernphysik, Technische Universit{\"a}t Darmstadt, D-64289 Darmstadt, Germany}

\author{T.~Friedrich}
%\email{friedrich@ikp.tu-darmstadt.de}
\affiliation{Institut f{\"u}r Kernphysik, Technische Universit{\"a}t Darmstadt, D-64289 Darmstadt, Germany}

\author{H.~L.~Harney}
%\email{hanns-ludwig.harney@mpi-hd.mpg.de}
\affiliation{Max-Planck-Institut f{\"u}r Kernphysik, D-69029 Heidelberg, Germany}

\author{M.~Miski-Oglu}
%\email{maksim@ikp.tu-darmstadt.de}
\affiliation{Institut f{\"u}r Kernphysik, Technische Universit{\"a}t Darmstadt, D-64289 Darmstadt, Germany}

\author{A.~Richter}
\email{richter@ikp.tu-darmstadt.de}
\affiliation{Institut f{\"u}r Kernphysik, Technische Universit{\"a}t Darmstadt, D-64289 Darmstadt, Germany}

\author{F.~Sch{\"a}fer}
%\email{schaefer@ikp.tu-darmstadt.de}
\affiliation{Institut f{\"u}r Kernphysik, Technische Universit{\"a}t Darmstadt, D-64289 Darmstadt, Germany}

\author{H.~A.~Weidenm{\"u}ller}
\affiliation{Max-Planck-Institut f{\"u}r Kernphysik, D-69029 Heidelberg, Germany}

\date{\today}

\begin{abstract}
Using reciprocity, we investigate the breaking of time-reversal (\T) symmetry
due to a ferrite embedded in a flat microwave billiard. Transmission spectra
of isolated single resonances are not sensitive to \T-violation whereas those
of pairs of nearly degenerate resonances do depend on the direction of time.
For their theoretical description a scattering matrix model from nuclear
physics is used. The \T-violating matrix elements of the effective Hamiltonian
for the microwave billiard with the embedded ferrite are determined
experimentally as functions of the magnetization of the ferrite.
\end{abstract}


\maketitle

%\section{Introduction}

We investigate time-reversal symmetry breaking (TRSB) in resonant scattering
through a flat microwave cavity. Such cavities, also known as "microwave
billiards," have long been used to simulate quantum
billiards~\cite{mw:billards, mw:chaos, richter:playing}, and interest has
mainly been focused on statistical properties of the
eigenvalues~\cite{so:guegoe, sto:trsbwg}, which have also been studied in
microwave networks~\cite{sirko:net}, and eigenfunctions~\cite{wu:ampfluc}, and
on dissipative transport~\cite{sch:tranprop}. We report on experiments where
time-reversal (\T) symmetry is broken by the magnetic and absorptive
properties of a piece of ferrite. Here, the two-state system is the simplest
one to show effects of TRSB. Thus, our investigations focus on pairs of nearly
degenerate resonances (doublets) and, as a test, also on isolated resonances
(singlets). We determine all relevant parameters of the system as functions of
the magnetization of the ferrite and thereby the properties of both, the
two-state system and the ferrite. Needless to say, our work is not related
directly to studies of TRSB in fundamental interactions. Indeed, we recall
that in the absence of dissipation all classical physics is \T invariant, and
that among the four fundamental interactions, only the weak force is known to
cause TRSB~\cite{kaon}. Rather, the microwave billiard with the ferrite inside
serves as a paradigmatic example of how to study and analyze TRSB in resonance
scattering experiments.

%\section{Experimental setup}

\begin{figure}[b]
	\centering
	\includegraphics[width=8cm]{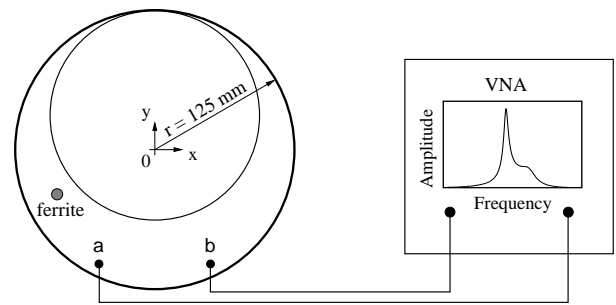}
	\caption{Scheme of the experimental setup (not to scale). The antennas
	connected to the vector network analyzer (VNA) were located at the positions
	$a$ and $b$. The inner circle is the copper disk introduced into the
	resonator to transform the circular into an annular billiard.}
	\label{fig:setup_scheme}
\end{figure}

The experiments were performed with a flat, cylindrical microwave resonator
made of copper. The resonator has a circular shape with a diameter of 250~mm
and a height of 5~mm (see Fig.~\ref{fig:setup_scheme}). An HP~8510C vector
network analyzer (VNA) coupled power into and out of the system through two
antennas, metal pins of 0.5~mm in diameter reaching about 2.5~mm into the
resonator. The microwave resonator with the antennas as entrance and exit
channels defines a scattering system~\cite{scattering}. The VNA measures the
relative phase and amplitude of the input and the output signals, thereby
providing the complex elements of the scattering matrix. Effects of the
connecting coaxial cables have been eliminated by a careful calibration of the
VNA. The TRSB is caused by a ferrite inside the resonator~\cite{so:guegoe,
sto:trsbwg, wu:ampfluc, sch:tranprop}. The ferrite is a calcium vanadium
garnet and has the shape of a cylinder, 5~mm high and 4~mm in
diameter~\cite{aft}. The material has a resonance line width of $\Delta H =
17.5~{\rm Oe}$ and a saturation magnetization of $4 \pi M_{\rm S} = 1859~{\rm
Oe}$. At the position of the ferrite a static magnetic field was created with
a NdFeB magnet, which was placed outside the resonator and connected to a
screw thread. Thus, magnetic field strengths of up to 90~mT (with an
uncertainty below 0.5~mT) were obtained at the ferrite.

%\section{T-violation with ferrites in theory and practice}

The spins within the ferrite precess with their Larmor frequency about the
static magnetic field; this introduces a chirality into the system. A
circularly polarized rf magnetic field interacts with the precessing spins if
it has the same rotational direction and frequency. The magnetic field
component of the rf field of the resonator can be split into two circularly
polarized fields rotating in opposite directions. Only one of these is
attenuated by the ferrite, this causes TRSB. In contradistinction to
Ref.~\cite{Fink}, \T invariance is experimentally tested by interchanging
input and output at the resonator. Thereby, the attenuated direction of
polarization is interchanged with the unaffected one. The TRSB effect is
strongest at the ferromagnetic resonance where the Larmor frequency of the
ferrite matches the eigenfrequency of the cavity.

%\section{Detailed balance and reciprocity}

To interpret our experiments, we use the scattering formalism for microwave
billiards of Ref.~\cite{scattering}. The element $S_{ab}$ of the scattering
matrix $S$ is the amplitude for the transition from an initial state (in the
connecting coaxial cables) $|b\rangle$ to a final state $|a\rangle$.  For a
resonator with two attached, pointlike antennas $S$ has dimension two.
\T invariance implies that $S$ is symmetric~\cite{henley:detbal}, i.\,e.
\begin{equation}
	S_{ab} = S_{ba} \ .
	\label{eqn:reciproci}
\end{equation}
Property~(\ref{eqn:reciproci}) is referred to as \emph{reciprocity}.
Experiments in nuclear physics~\cite{nuclear:detbal} have tested the weaker
principle of detailed balance~\cite{henley:detbal} which implies $|S_{ab}|^2 =
|S_{ba}|^2$. In the present experiment we test the stronger statement of
reciprocity.

How does the absorption of the circularly polarized rf field lead to a
violation of Eq.~(\ref{eqn:reciproci})?  We write $S$ in the form
\begin{equation}
	S_{ab}(\omega) = \delta_{ab} - 2\pi\, i\, 
		\langle a | W^\dagger\, (\omega-H^{\rm eff})^{-1}\, W | b \rangle\, ,
	\label{eqn:Sab}
\end{equation}
originally derived in the context of nuclear physics in Refs.~\cite{mahaux:66,
mahaux:69}. Here $\omega/(2\pi)$ is the frequency of the rf field. The matrix
$W$ describes the coupling of the waves in the coaxial cables ($\vert a
\rangle$, $\vert b \rangle$) with the resonator state(s). Since that coupling
is \T invariant, $W$ can be chosen real. For a single resonance with resonator
state $\vert 1 \rangle$, we deal with the two matrix elements $\langle 1 \vert
W \vert a \rangle$ and $\langle 1 \vert W \vert b \rangle$. For a doublet with
the two resonator states $\vert 1 \rangle$ and $\vert 2 \rangle$, $W$ can be
written as
\begin{equation}
	W | a \rangle = N_a
		\left(
			\begin{array}{c} \cos \alpha \\ \sin \alpha \end{array}
		\right),\quad 
	W | b \rangle = N_b
		\left(
			\begin{array}{c} \cos \beta \\ \sin \beta \end{array}
		\right)\, .
	\label{eqn:VaVb}
\end{equation}
The resonator with the ferrite is represented by the effective Hamiltonian
$H^{\rm eff}$. For a doublet of resonances, $H^{\rm eff}$ has dimension two.
The frequencies $\omega_{\mu}$ with $\mu = 1,2$ of the resonances appear in
the diagonal elements. In addition, $H^{\rm eff}$ contains the matrix (see
Eq.~(4.2.20b) of Ref.~\cite{mahaux:69})
\begin{equation}
	F_{\mu \nu}(\bar{\omega}) = \sum_{i = a,b,f} \int_{-\infty}^{\infty} {\rm d}\omega'
		\frac{W_{\mu i}(\omega') W_{\nu i}^\ast(\omega')}
		     {\bar{\omega}^+ - \omega'}\, ,
	\label{eqn:Heff}
\end{equation}
where $\bar{\omega}^+ = \bar{\omega} + i\,\varepsilon$ is the angular
frequency $\bar{\omega}$ shifted infinitesimally to complex values with
$\varepsilon > 0$. In the sum on the right-hand side of Eq.~(\ref{eqn:Heff}),
the terms with $i = a,b$ stand for the couplings to the antennas with matrix
elements $W_{\mu i}$ as in Eq.~(\ref{eqn:VaVb}). As these are real, the
integration over $W_{\mu i}(\omega') W_{\nu i}^\ast(\omega')$ yields for $i =
a, b$ a complex symmetric matrix. However, because of TRSB the matrix elements
$W_{\mu f}$ that couple the resonator states $|\nu\rangle,\ \nu=1, 2$, to the
ferromagnetic resonance in the ferrite, cannot be chosen real. Hence the
integration over $W_{\mu f}(\omega') W_{\nu f}^\ast(\omega')$ leads to both, a
complex symmetric matrix and a complex antisymmetric matrix $H^{\rm a}$ with
\begin{equation}
	H^{\rm a} = 
		\left( 
			\begin{array}{cc} 0 & H_{12}^{\rm a} \\ -H^{\rm a}_{12} & 0 \end{array} 
		\right) \ .
	\label{eqn:Hamatrix}
\end{equation}
The matrix $H^{\rm a}$ is the hallmark of TRSB in $H^{\rm eff}$. Since $H^{\rm
a}$ is antisymmetric, TRSB cannot be observed for a single isolated resonance.
In the case of a pair of resonances, it is useful to choose the doublet
spacing as small as possible to enhance TRSB. We observe that $H^{\rm a}$ does
not depend on the choice of the resonator basis (it is invariant under
orthogonal transformations).

%\section{Isolated resonances}

\begin{figure}[b]
	\centering
	\includegraphics{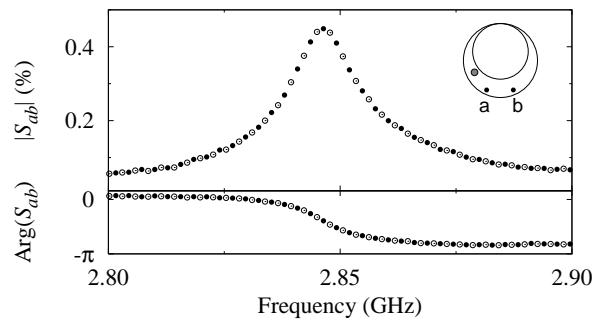}
	\caption{A singlet in the annular billiard. For an external field of
	119.3~mT we show $S_{ab}$ (open circles) and $S_{ba}$ (filled circles). Both
	amplitudes and phases coincide and reciprocity holds. The statistical errors
	of the data are smaller than the symbols.}
	\label{fig:ann_sing}
\end{figure}

As a test of these arguments we first discuss isolated resonances (singlets).
It is well known that in circular billiards resonances with vanishing angular
momentum are already singlets~\cite{mw:chaos}. However, our billiard provides
only two truly isolated singlets. Therefore, we suppressed the existing
degeneracies by breaking the rotational symmetry. We have introduced a copper
disk (187.5~mm diameter, 5~mm height) into the resonator, see
Fig.~\ref{fig:setup_scheme}. The resulting "annular billiard" is fully
chaotic~\cite{dembo:annular}. In the presence of the ferrite, its spectrum
contains several well isolated resonances (widths $\approx$ 14~MHz, spacings
$\approx$ 300~MHz). We have studied eight singlets. For the singlet at
frequency $f=2.84~{\rm GHz}$, we show in Fig.~\ref{fig:ann_sing} both $S_{ab}$
and $S_{ba}$. Within the errors due to noise the complex function $S_{ab}$
agrees with $S_{ba}$; i.\,e.\ reciprocity holds despite the presence of the
ferrite. The same statement applies to the other seven singlets. This shows
that the coupling to the leads (see Eq.~(\ref{eqn:Sab})) is indeed
\T invariant.

%\section{Doublet resonances}

\begin{figure}[tb]
	\centering \includegraphics{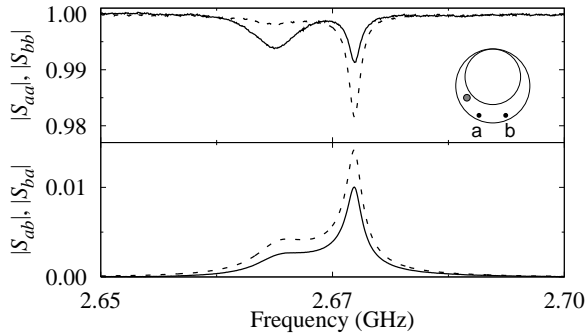}
	\caption{The doublet at 2.67~GHz in the circular billiard with an external
	magnetic field of 36.0~mT. The upper part shows the absolute values of
	$S_{aa}$ (solid) and $S_{bb}$ (dashed), the lower one those of $S_{ab}$
	(solid) and $S_{ba}$ (dashed). The uncertainties of $S_{ab}$ and $S_{ba}$
	are about $5\times10^{-4}$. Reciprocity is clearly violated.}
	\label{fig:circle_doub}
\end{figure}

We used the circular billiard (see Fig.~\ref{fig:setup_scheme}) with its
numerous degeneracies to study isolated doublets of resonances. The ferrite
destroys the rotational symmetry and, thus, partly lifts the degeneracies. We
found four isolated doublets at 2.43~GHz, 2.67~GHz, 2.89~GHz and 3.2~GHz. The
influence of the magnetized ferrite on the resonance shape of the second
doublet is illustrated in Fig.~\ref{fig:circle_doub}. In contrast to the case
of singlets, the direction of transmission is now important; reciprocity is
violated. This was also observed for the first and the third doublet, but not
for the fourth. A simulation of the electromagnetic field patterns inside the
resonator~\cite{mws} showed that for one of the two states of the fourth
doublet, the magnetic field vanished at the location of the ferrite, so that
the ferrite interacted with just one state and $H^{\rm a} = 0$. Moving the
ferrite to a position where it interacted with both states, resulted in a
violation of reciprocity. This confirms the importance of coupling a pair of
states for observing TRSB.

%\section{Scattering theory}

We measured all four $S$-matrix elements ($S_{aa}$, $S_{ab}$, $S_{ba}$,
$S_{bb}$) of the system, each at 15 settings of the external magnetic field.
From these data, it is possible to determine $H^{\rm a}_{12}$. Indeed,
Eq.~(\ref{eqn:Sab}) has twelve real parameters: $\alpha$, $\beta$, $N_a$,
$N_b$ and the four complex matrix elements of $H^{\rm eff}$. We assume that
$\alpha$, $\beta$, $N_a$, $N_b$, $W_{\mu a}$, $W_{\mu b}$ are independent of
the external magnetic field and, within the range of frequencies of the
doublet, of $\omega$.

\begin{figure}[tb]
	\centering
	\includegraphics[width=8cm]{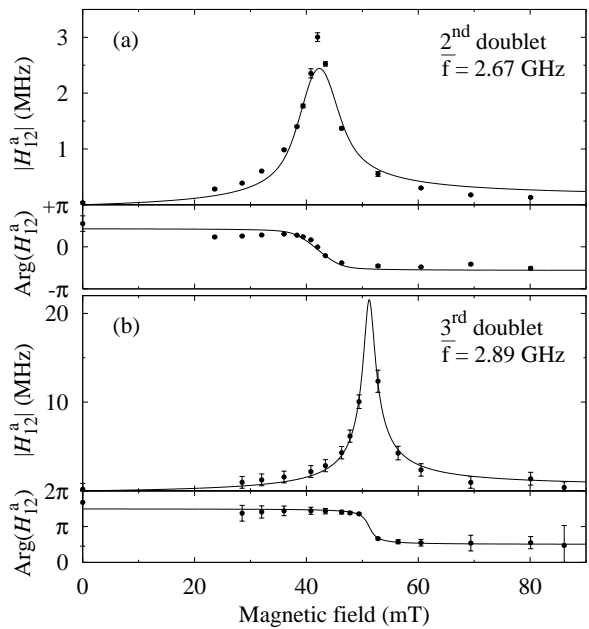}
	\caption{The \T violating matrix element $H^{\rm a}_{12}$ for (a) the second
	doublet at 2.67~GHz (upper panel) and (b) the third doublet at 2.9~GHz
	(lower panel). The error bars include the systematic, statistical, and
	numerical errors. The line shape of the ferromagnetic resonance for the
	second doublet (upper panel) has been convoluted with a Gaussian
	distribution of magnetization with width 15.5~Oe, see
	Eq.~(\ref{eqn:HaConv}).}
	\label{fig:hasym}
\end{figure}

For the second and third doublet, the values deduced for $H^{\rm a}_{12}$ are
shown as dots in Fig.~\ref{fig:hasym}. Care was taken to rule out any TRSB
effects feigned by the VNA. The data points show a clear resonancelike
structure: As a function of the external magnetic field, $H^{\rm a}_{12}$ goes
through a maximum; at the same time the phase of $H^{\rm a}_{12}$ decreases by
about $\pi$. We now show that this behavior is intimately connected with the
ferromagnetic resonance.

%\section{T-violating matrix element theory}

We use Eq.~(\ref{eqn:Heff}) to express $H^{a}_{12}$ in terms of the matrix
elements $W_{\nu f}(\omega')$. As explained above, the ferrite is coupled to
only one of the two circular polarizations of the rf magnetic field in the
resonator. The unitary matrix
\begin{equation}
	U = \frac{1}{\sqrt{2}} 
		\left(
    	\begin{array}{rr} 1 & -i \\ 1 & i \end{array}
		\right)
	\label{eqn:U}
\end{equation}
induces a transformation from the real resonator basis states $\{|1\rangle,
|2\rangle\}$ with matrix elements $W$ to circular basis states with matrix
elements $\tilde{W}$ so that $W_{\nu f}(\omega') = \sum_{\xi = 1}^2 U^*_{\xi
\nu} \tilde{W}_{\xi f}(\omega')$. By definition, one of the $\tilde{W}_{\xi
f}$ ($\tilde{W}_{2 f}$, say) vanishes so that $W_{1 f}(\omega) W^\ast_{2
f}(\omega) = |\tilde{W}_{1 f}(\omega)|^2 / (2 i)$. Moreover, $\tilde{W}_{1 f}$
is proportional~\cite{lax:ferrit} to the magnetic susceptibility
\begin{equation}
	\tilde{W}_{1 f}(\omega') 
		\propto \chi(\omega') 
		= \frac{\omega_M}{\omega_0(B) - \omega' - i/T}\, ,
	\label{eqn:chi}
\end{equation}
where $T(\Delta H)=\left( \pi\, \gamma\, \Delta H \right)^{-1}$ is the
relaxation time, $\omega_M = 2\, \pi\, \gamma\, 4\pi M_S$ with the
gyromagnetic ratio $\gamma = 2.8\ {\rm MHz}/{\rm Oe}$ and
\begin{equation}
	\omega_0(B) = 2\pi \left(1.7~{\rm GHz} + 0.0237~{\rm GHz}/{\rm mT}~B\right).
	\label{eqn:omega0B}
\end{equation}
The relation between the strength of the external magnetic field $B$ and the
frequency of the ferromagnetic resonance $\omega_0(B)$ has been established by
prior measurements with the ferrite in a waveguide.

Collecting Eqs.~(\ref{eqn:Heff})--(\ref{eqn:omega0B}), we find
\begin{equation}
	H_{12}^a(B) 
		= \frac{i \pi}{2}\, \lambda\, B\, T\, 
		  \frac{\omega_M^2}{\omega_0(B) - \bar{\omega} - i/T}\, .
	\label{eqn:Ha}
\end{equation}
The proportionality stated in Eq.~(\ref{eqn:chi}) has here been expressed in
terms of a linear dependence on $B$ and on a coupling strength parameter
$\lambda$. Equation~(\ref{eqn:Ha}) depends on two free parameters,
$\bar{\omega}$ and $\lambda$. The parameter $\bar{\omega}$ averages the TRSB
effect of the ferrite over a doublet and is expected to be close to the
resonance frequencies of the doublet. Equations~(\ref{eqn:omega0B}) and
(\ref{eqn:Ha}) imply that the phase of $H^{\rm a}_{12}$ decreases with
increasing~$B$, independently of the choice of the phase of the matrix
element~(\ref{eqn:chi}). This is in agreement with the data.

%\section{Experimental results}

In Fig.~\ref{fig:hasym}b, we show for the third doublet the measured values of
$H^{\rm a}_{12}$ and the function~(\ref{eqn:Ha}). The agreement is very good.
We cannot obtain data points at the maximum of the ferromagnetic resonance
since there the susceptibility~(\ref{eqn:chi}) varies appreciably over the
frequency range of the doublet. The theory of
Eqs.~(\ref{eqn:Heff})--(\ref{eqn:Ha}) is not valid in that regime. For the
second doublet shown in Fig.~\ref{fig:hasym}a, the resonance is too broad to
be explicable by a ferromagnetic resonance width of $\Delta H = 17.5~{\rm
Oe}$. This is because Eqs.~(\ref{eqn:chi}) to (\ref{eqn:omega0B}) apply to a
completely saturated ferrite only. Saturation is achievable only with a strong
magnetic field, however. To account for the incomplete saturation we assume
that the magnetization has a Gaussian distribution with r.m.s. value $\sigma$
and phase $\alpha$, $\Psi(\omega) = e^{i \alpha} e^{-(1/2)( \omega /
\sigma)^2}/(\sigma\sqrt{2 \pi}).$ The convolution with $H^{\rm a}_{12}$ of
Eq.~(\ref{eqn:Ha}),
\begin{equation}
	\tilde{H}_{12}^a(B) = \int_{-\infty}^\infty {\rm d} \omega'\,
		H_{12}^a(\omega')\, \Psi[\omega_0(B) - \omega']\, ,
	\label{eqn:HaConv}
\end{equation}
yields agreement with the experimental data [see Fig.~\ref{fig:hasym}a]. The
numerical values of the parameters in Eq.~(\ref{eqn:Ha}) are given in
Table~\ref{tab:params}. We note that $\lambda$ depends on the properties of the
ferrite as well as on the values of the wave functions of the doublet states
at the position of the ferrite.

\begin{table}[tb]
	\centering
	\caption{Parameters of Eq.~(\ref{eqn:HaConv}) for the first three doublets.
	Doublet No.\ 3 is not convoluted; hence $\sigma$ and $\alpha$ are not defined
	in this case.}
	\begin{tabular*}{\columnwidth}{c@{\extracolsep{\fill}}crrrrc}
		\hline 
		\hline 
		& \# & $\sigma$ (Oe) & $\bar{\omega}/2\pi$ (GHz) & $\lambda$ (mT$^{-1}$) & $\alpha$ (deg) & \\ 
		\hline 
		& 1 & $42.1\pm9.3$ & $2.427\pm0.037$ & $35.7\pm4.6$ & $ -5\pm8$ & \\
		& 2 & $15.5\pm3.3$ & $2.696\pm0.011$ & $10.8\pm1.0$ & $168\pm7$ & \\
		& 3 &  $\cdots$    & $2.914\pm0.003$ & $37.3\pm1.6$ &  $\cdots$ & \\
		\hline 
		\hline
	\end{tabular*}
	\label{tab:params}
\end{table}

%\section{Conclusion}

In summary, we have investigated TRSB in singlets and doublets in
ferrite-loaded cavities. To the best of our knowledge this is the first time
that the principle of reciprocity has been investigated in microwave
resonators with broken time-reversal symmetry. While TRSB cannot be detected
at isolated resonances, we have shown that reciprocity is violated for
resonance doublets. We extracted the \T violating matrix element of the
effective Hamiltonian which describes the resonator plus the ferrite. That
matrix element can be modelled in terms of a ferromagnetic resonance due to
the coupling of the rf magnetic field with the ferrite. We plan to use these
insights to study intensity fluctuations at high excitation energies ("Ericson
fluctuations"~\cite{nuclear:detbal}) as well as exceptional
points~\cite{dembo:ep, hlh:epTRSB} in a TRSB regime.

\begin{acknowledgments}
We thank H.\ Benner and U.\ Hoeppe for discussions on the ferromagnetic
resonance. F.~S. is grateful for the financial support from GRK~410. This work
was supported by the DFG within SFB~634.
\end{acknowledgments}

\end{document}